\documentclass[12pt]{article}
\input{psfig}
 \usepackage{pictex}
 \usepackage{epsfig}
 \usepackage{pstricks}
 \usepackage{pst-coil}

\def\a{\alpha}
\def\b{\beta}

\def\d{\delta}
\def\e{\epsilon}
\def\f{\phi}
\def\g{\gamma}
\def\h{\eta}

\def\l{\lambda}
\def\m{\mu}
\def\n{\nu}
\def\o{\omega}
\def\p{\pi}

\def\r{\rho}

\def\D{\Delta}
\def\F{\Phi}
\def\G{\Gamma}

\def\P{\Pi}
\def\Q{\Theta}
\def\S{\Sigma}

\def\cl{{\cal L}}
\def\cm{{\cal M}}

\def\cp{{\cal P}}

\def\slpa{\slash{\pa}}                            
\def\bo{{\raise.15ex\hbox{\large$\Box$}}}               
\def\pa{\partial}                                       
\def\pr{\prod}                                          
\def\face{{\raise.2ex\hbox{$\displaystyle \bigodot$}\mskip-2.2mu \llap {$\ddot
        \smile$}}}                                      
\def\dg{\dagger}                                     
\def\wt#1{\widetilde{#1}}                    
\def\beq{\begin{equation}}
\def\eeq{\end{equation}}
\def\bea{\begin{eqnarray}}
\def\eea{\end{eqnarray}}
\def\NO{\nonumber}

\def\slash#1{\rlap{\hbox{$\mskip 1 mu /$}}#1}      
\def\wt#1{\widetilde{#1}}                    
\def\Bar#1{\overline{#1}}                       
\def\leftrightarrowfill{$\mathsurround=0pt \mathord\leftarrow \mkern-6mu
        \cleaders\hbox{$\mkern-2mu \mathord- \mkern-2mu$}\hfill
        \mkern-6mu \mathord\rightarrow$}       
\def\dvec#1{\vbox{\ialign{##\crcr
        \leftrightarrowfill\crcr\noalign{\kern-1pt\nointerlineskip}
        $\hfil\displaystyle{#1}\hfil$\crcr}}}           
\def\der#1{{\pa \over \pa {#1}}}                

\def\pl#1#2#3{Phys.~Lett.~{\bf B {#1}} (19{#2}) #3}
\def\np#1#2#3{Nucl.~Phys.~{\bf B {#1}} (19{#2}) #3}
\def\prl#1#2#3{Phys.~Rev.~Lett.~{\bf #1} (19{#2}) #3}
\def\pr#1#2#3{Phys.~Rev.~{\bf D {#1}} (19{#2}) #3}

\catcode`@=11
\def\@citex[#1]#2{\if@filesw\immediate\write\@auxout{\string\citation{#2}}\fi
  \def\@citea{}\@cite{\@for\@citeb:=#2\do
    {\@citea\def\@citea{,\penalty\@m}\@ifundefined
      {b@\@citeb}{{\bf ?}\@warning
       {Citation `\@citeb' on page \thepage \space undefined}}%
\hbox{\csname b@\@citeb\endcsname}}}{#1}}

\def\citer{\@ifnextchar [{\@tempswatrue\@citexr}{\@tempswafalse\@citexr[]}}
\def\@citexr[#1]#2{\scriptsize 
  \if@filesw\immediate\write\@auxout{\string\citation{#2}}\fi
  \def\@citea{}\@cite{\@for\@citeb:=#2\do
    {\@citea\def\@citea{-\penalty\@m}\@ifundefined
       {b@\@citeb}{{\bf ?}\@warning
       {Citation `\@citeb' on page \thepage \space undefined}}%
\hbox{\csname b@\@citeb\endcsname}}}{#1}\normalsize}
\catcode`@=12

\catcode`\@=11
\long\def\@makefntext#1{ 
\protect\noindent \hbox to 3.2pt {\hskip-.9pt  
$^{{\ninerm\@thefnmark}}$\hfil}#1\hfill} 

 \def\@makefnmark{\hbox to 0pt{$^{\@thefnmark}$\hss}}  
        
\def\ps@myheadings{\let\@mkboth\@gobbletwo
\def\@oddhead{\hbox{} 
\rightmark\hfil\ninerm\thepage}   
\def\@oddfoot{}\def\@evenhead{\ninerm\thepage\hfil 
\leftmark\hbox{}}\def\@evenfoot{}
\def\sectionmark##1{}\def\subsectionmark##1{}}

\begin{document}

\date{}
\title{
{\normalsize\rm DESY 00-056}\hfill{\mbox{}}\\
{\normalsize\rm April 2000}\hfill{\mbox{}}\vspace*{2cm}\\
{\bf QUANTUM MECHANICS OF BARYOGENESIS}
}
\author{W.~Buchm\"uller and S.~Fredenhagen  \\
\vspace{3.0\baselineskip}                                               
{\normalsize\it Deutsches Elektronen-Synchrotron DESY, 22603 Hamburg, Germany}
\vspace*{1cm}                     
}        

\maketitle

\thispagestyle{empty}

\begin{abstract}
\noindent
The cosmological baryon asymmetry can be explained as remnant of heavy
Majorana neutrino decays in the early universe. We study this 
out-of-equilibrium process by means of Kadanoff-Baym equations which
are solved in a perturbative expansion. To leading order the problem is
reduced to solving a set of Boltzmann equations for distribution functions. 
\end{abstract}

\newpage
The generation of the cosmological matter-antimatter asymmetry in an expanding
universe requires baryon number violation, $C$ and $C\!P$ violation, and a
deviation from thermal equilibrium \cite{sa67}. The classic mechanism which
realizes these conditions is the decay of weakly interacting massive particles
in a thermal bath \cite{yo78}. Particularly successful is the leptogenesis 
scenario where the decaying particles are heavy Majorana neutrinos 
\cite{fy86}. The resulting baryon asymmetry is entirely determined by 
neutrino properties. The observed order of magnitude can be naturally 
explained without any fine tuning of parameters and in accord
with present experimental indications for neutrino masses \cite{bp96}.

The generation of a baryon asymmetry is an out-of-equilibrium process 
which is generally treated by means of Boltzmann equations. A thorough
discription of the basic ideas can be found in \cite{kw80}. Some subtleties 
have recently been discussed in \cite{bf00}. A shortcoming of this
approach is that the Boltzmann equations are classical equations for the time 
evolution of phase space distribution functions. On the contrary, the
involved collision terms are $S$-matrix elements which involve
quantum interferences of different amplitudes in a crucial manner. Clearly,
a full quantum mechanical treatment is highly desirable. It is also
required in order to justify the use of Boltzmann equations and to determine 
the size of corrections.

All information about the time evolution of a system is contained in the
time dependence of its Green functions \cite{kb62,ke64},
which can be determined by means of Dyson-Schwinger equations. Originally
these techniques were developed for non-relativistic many-body problems. More 
recently, they have also been applied to transport phenomena in nuclear matter
\cite{mh94}, the electroweak plasma \cite{ri98,jkp00} and the QCD plasma 
\cite{bi99}. Alternatively, one may study the time evolution of density 
matrices \cite{jmy98,abv98}. In the following we shall investigate  
non-equilibrium Green functions which are relevant for leptogenesis. We shall 
construct a perturbative solution of the corresponding Kadanoff-Baym equations
which, to leading order, turn out to be equivalent to a set of Boltzmann 
equations. Higher-order corrections can then be systematically evaluated.

Consider now the standard model with three additional right-handed neutrinos
whose interactions are described by the lagrangian,
\beq
\cl = \Bar{l}_L \wt{\f}\l^* \n_R - {1\over 2}\Bar{\n}_R^cM\n_R + h.c.
\eeq
Here $l_L$ and $\f$ denote lepton and Higgs doublets, respectively.
We shall restrict our discussion to the case of hierarchical Majorana neutrino
masses, $M_1 \ll M_2,M_3$. The baryon asymmetry will then be determined by
the $C\!P$ violating decays of the lightest Majorana neutrino
$N_1=\n_{R1}+\n_{R1}^c\equiv N$,
\beq\label{gcp} 
\G(N\rightarrow l\phi) = {1\over 2}(1+\e)\G\;,\quad
\G(N\rightarrow \bar{l}\bar{\phi})={1\over 2}(1-\e)\G\;.
\eeq
Here $\G$ is the total decay width and the parameter $\e \ll 1$ measures the 
amount of $C\!P$ violation. The generation of the baryon asymmetry takes
place at a temperature $T\sim M_1\equiv M \ll M_2,M_3$. It is therefore
convenient to describe the system by an effective lagrangian where the two
heavier neutrinos have been integrated out,
\bea\label{lint}
\cl &=& \Bar{l}_{Li} \wt{\f}\l^*_{i1} N + 
      N^T \l_{i1} C l_{Li}\f - {1\over 2}M N^T C N \NO\\
&& + {1\over 2}\h_{ij} l_{Li}^T\f\ C\ l_{Lj}\f
   + {1\over 2}\h^*_{ij} \Bar{l}_{Li}\wt{\f}\ C\ \Bar{l}_{Lj}^T\wt{\f}\;,
\eea
with
\beq
\h_{ij}=\sum_{k=2}^3\l_{ik}{1\over M_k}\l^T_{kj}\;.
\eeq

For leptogenesis one has to consider the phase space distributions for heavy 
neutrinos ($f_N$), leptons ($f_l$), anti-leptons ($f_{\bar{l}}$), 
Higgs ($f_\f$) and anti-Higgs bosons ($f_{\bar{\f}}$). The generation 
of the lepton asymmetry is a process close to equilibrium. Hence one can 
linearize the Boltzmann equations in 
the deviations from the equilibrium distributions. Due to the interactions 
in (\ref{lint}) one has $\d f_l = -\d f_{\bar{l}} = \d f_\f 
= -\d f_{\bar{\f}}$. The Boltzmann equation for the Majorana neutrino reads
\bea
&&g_N\der t \d f_N(t,p) = \label{bon}\\
&&\quad - g_N\der t f_N(p) 
-{1\over 2E} \int d\F_{\bar{1}\bar{2}}(p)
\d f_N(t,p) \left(|\cm(N\rightarrow l\f)|^2 + 
  |\cm(N\rightarrow \bar{l}\bar{\f})|^2\right). \NO
\eea
For the lepton doublets one obtains
\bea
&&2g_l\der t \d f_l(t,k) = \NO\\
&&\quad {1\over 2k} \int d\F_{\bar{1}2}(k)\
\e\ \d f_N(t,p_1) \left(|\cm(N\rightarrow l\f)|^2 + 
  |\cm(N\rightarrow \bar{l}\bar{\f})|^2\right)\NO\\
&&\quad-{1\over 2k}\int d\F_{1\bar{2}}(k) \left(\d f_l(t,k)f_\f(p_1)
                    +f_l(k)\d f_\f(t,p_1)\right)\NO\\
&&\quad\hspace{2.5cm}\times\left(|\cm(l\f\rightarrow N)|^2
   +|\cm(\bar{l}\bar{\f}\rightarrow N)|^2\right)\NO\\
&&\quad-{1\over 2k} \int d\F_{1\bar{2}\bar{3}}(k)
\left(\d f_l(t,k)f_\f(p_1)+f_l(k)\d f_\f(t,p_1)\right)\NO\\
&&\quad\hspace{2.5cm}\times\left(|\cm(l\f\rightarrow \bar{l}\bar{\f})|^2
+|\cm(\bar{l}\bar{\f}\rightarrow l\f)|^2\right)\NO\\
&&\quad-{1\over 2k} \int d\F_{\bar{1}\bar{2}3}(k)
\left(\d f_l(t,p_1)f_\f(p_2)+f_l(p_1)\d f_\f(t,p_2)\right)\NO\\
&&\quad\hspace{2.5cm}\times\left(|\cm(l\f\rightarrow \bar{l}\bar{\f})|^2
+|\cm(\bar{l}\bar{\f}\rightarrow l\f)|^2\right)\NO\\
&&\quad-{1\over 4k} \int d\F_{1\bar{2}\bar{3}}(k)
\left(\d f_l(t,k)f_l(p_1)+f_l(k)\d f_l(t,p_1)\right)\NO\\
&&\quad\hspace{2.5cm}\times\left(|\cm(ll\rightarrow \bar{\f}\bar{\f})|^2
   +|\cm(\bar{l}\ \bar{l}\rightarrow \f\f)|^2\right)\NO\\
&&\quad-{1\over 2k} \int d\F_{\bar{1}\bar{2}3}(k)\d f_\f(t,p_1)f_\f(p_2)
\left(|\cm(\f\f\rightarrow \bar{l}\ \bar{l})|^2+
|\cm(\bar{\f}\bar{\f}\rightarrow ll)|^2\right)\;.\label{bol}
\eea
Here 
\beq
d\Phi_{1\ldots \bar{n}\ldots}(p)=
{d^3p_1\over (2\pi)^3 2E_1}\ldots {d^3p_{\bar{n}}\over (2\pi)^3 2E_{\bar{n}}}
\ldots(2\pi)^4 \d^4(p+p_1+\ldots-p_{\bar{n}}-\ldots)\;, 
\eeq
\beq
f_i(p)=\exp{(-\b E_i(p))}\;,
\eeq
denote phase space integrations and distribution functions, respectively.
The temperature $T=1/\b$, and $\cm(\ldots)$ is the matrix element
of the indicated process. $g_N=2$ and $g_l=g_{\bar{l}}=6$ are the number of 
`internal' degrees of freedom for the Majorana neutrino and the lepton 
doublets for three generations, respectively. For simplicity we have assumed 
small number densities
so that we can use Boltzmann distribution functions for bosons and fermions
and also neglect distribution functions for particles in the final state. The
effect of the Hubble expansion is included by introducing the `covariant'
derivative $\pa/\pa t \rightarrow \pa/\pa t - H p \pa/\pa p$. Integration over
momenta then yields the more familiar form of the Boltzmann equations for the 
number densities. Eq.~(\ref{bon}) describes the decay of the heavy Majorana 
neutrinos. Note, that also the equilibrium distributions are time dependent 
since the temperature varies with time. For massless particles the 
distribution functions are constant with respect to the `covariant' 
time derivative. The first term in eq.~(\ref{bol}) drives the generation of a 
lepton asymmetry; the remaining terms tend to wash out an existing asymmetry. 
Eqs.~(\ref{bon}) and (\ref{bol}) determine $\d f_N$ and $\d f_l$ as function
of time. We have only kept the interactions given by the lagrangian 
(\ref{lint}). A complete discussion can be found in \cite{pl97}.\\

\noindent\textbf{Green functions near thermal equilibrium}\\

The time evolution of an arbitrary multi-particle lepton-Higgs system can be 
studied by means of the Green functions of lepton and Higgs fields. For the 
heavy Majorana neutrino one has
\beq
iG_{\a\b}(x_1,x_2)=\mbox{Tr}\left(\r T N_\a(x_1)N_\b(x_2)\right)\;,
\eeq
where $T$ denotes the time ordering, $\r$ is the density matrix of the system,
the trace extends over all states, and the time coordinates $t_1$ and $t_2$
lie on an appropriately chosen contour $C$ in the complex plane \cite{lb96}. 
$G(x_1,x_2)$ can be written as a sum of two parts,
\beq\label{gn}
G(x_1,x_2)=\Q(t_1-t_2) G^>(x_1,x_2)+\Q(t_2-t_1) G^<(x_1,x_2)\; ,
\eeq  
where
\beq
iG^>(x_1,x_2)_{\a\b}=\mbox{Tr}\left(\r N_\a(x_1)N_\b(x_2)\right),\;
iG^<(x_1,x_2)_{\a\b}=-\mbox{Tr}\left(\r N_\b(x_2)N_\a(x_1)\right)\ .
\eeq
The `time ordering' in eq.~(\ref{gn}) is along the contour $C$.

For a system in thermal equilibrium at a temperature $T=1/\b$ the density 
matrix is $\r=\exp{(-\b H)}$, where $H$ is the Hamilton operator. In this
case the Green function only depends on the difference of coordinates and it 
is convenient to introduce the Fourier transform,
\beq
G(p) = \int d^4x e^{ipx} G(x)\; .
\eeq
The contour $C$ can be chosen as a sum of two branches, $C=C_1\cup C_2$, which
lie above and below the real axis. The time coordinates are real 
and associated with one of the two branches. Correspondingly, the Green 
function becomes a $2\times 2$ matrix,
\bea\label{gmatrix}
G(p)=\left(
\begin{array}{cc}
G^{11}(p) & G^{12}(p)  \\[1ex]
G^{21}(p) & G^{22}(p) 
\end{array}
\right)\; . 
\eea
The off-diagonal terms are given by
\beq
G^{12}(p)=G^<(p)\;, \quad G^{21}(p)=G^>(p)\;.
\eeq
The diagonal terms of the matrix (\ref{gmatrix}) are the familiar causal and 
anti-causal Green functions. The functions $G^>(p)$ and $G^<(p)$ satisfy the
KMS-condition,
\beq
G^<(p)=-e^{-\b p_0}G^>(p)\;,
\eeq
and the free Green functions are explicitly given by
\bea
iG^>(p)&=&\left(\Q(p_0)
                -\Q(p_0)f_N(E)-\Q(-p_0)f_{\bar{N}}(E)\right)\r_N(p)\;,\\ 
iG^<(p)&=&\left(\Q(-p_0)  
                 -\Q(p_0)f_N(E)-\Q(-p_0)f_{\bar{N}}(E)\right)\r_N(p)\;,
\eea
with the spectral density
\beq\label{nspectral}
\r_N(p) = 2\p (\slash p + M)C^{-1}\d(p^2-M^2)\;,
\eeq 
and the Fermi-Dirac distribution functions
\beq\label{fn}
f_N(E)=f_{\bar{N}}(E)={1\over e^{\b E} + 1}\ , \quad E=\sqrt{M^2+p^2}\; .
\eeq
Since $N(x)$ is a Majorana field one has $f_N=f_{\bar{N}}$, and in the
spectral density (\ref{nspectral}) the charge conjugation matrix $C$ occurs. 
In the following we shall also need the retarded and advanced Green functions,
\beq
G^{\pm}(x)=\pm\Q(\pm x^0) \left(G^>(x)-G^<(x)\right)\; ,
\eeq
which can be written as sum of an on-shell and an off-shell contribution,
\beq\label{onoff}
G^{\pm}(p)=\pm {1\over 2}\left(G^>(p)-G^<(p)\right)
+{1\over 2\p i} \cp \int d\o' {G^>(x,\o',\vec{p})-G^<(x,\o',\vec{p})
                    \over \o-\o'}\; .
\eeq

The Green functions for the lepton doublets and for the Higgs doublet,
\beq
iS(x_1,x_2)_{\a\b}\d^a_b=
\mbox{Tr}\left(\r T l_{\a}^a(x_1)\bar{l}_{b\b}(x_2)\right) ,
\; i\D(x_1,x_2)\d^a_b=\mbox{Tr}\left(\r T \phi^a(x_1)\phi_b^*(x_2)\right),
\eeq
have the same structure as $G(x_1,x_2)$. The corresponding equations for
$S(p)$ are obtained from eqs.~(\ref{gn})-(\ref{fn}) by replacing the  
spectral density $\r_N(p)$ by 
\beq
\r_l(p)=2\p P_L \slash p \d(p^2-M^2)\;, \quad P_L={1-\g_5\over 2}\;,
\eeq
and the distribution functions $f_N(E)$ and $f_{\bar{N}}(E)$ by
\beq
f_l(E,\m_l)=f_{\bar{l}}(E,-\m_l)={1\over e^{\b (E-\m_l)} + 1}\ ,
\quad E=|\vec{p}|\;,
\eeq
where $\m_l$ is the lepton chemical potential. For the Higgs field one has
\beq
\r_{\phi}(p)=2\p \d(p^2-M^2)\;,
\eeq
\beq 
f_{\f}(E,\m_\f)=f_{\bar{\f}}(E,-\m_\f)={1\over e^{\b (E-\m_\f)} - 1}\ ,
\quad E=|\vec{p}|\;.
\eeq

We are considering a process close to equilibrium. This suggests that the
corresponding deviations of the Green functions may be obtained from the
equilibrium Green functions by a small change of the distribution functions,
\beq
i\d G(x,p) = -\d f_N(x,p)\r_N(p)\;, \label{devn}
\eeq
\beq
i\d S(x,k) = -\e(k_0) \d f_l(x,k)\r_l(k)\;,
\quad i\d \D(x,q) = -\e(q_0) \d f_{\f}(x,q)\r_{\f}(q)\;. \label{devl}
\eeq
Here we have used that due to the interactions given in (\ref{lint})
$\d f_l = -\d f_{\bar{l}} = \d f_\f = -\d f_{\bar{\f}}$.\\

\noindent\textbf{Kadanoff-Baym equations}\\

The Green functions for the heavy neutrino and the leptons satisfy
Dyson-Schwinger equations,
\bea
C(i\slpa_1-M)G(x_1,x_2)&=&\d(x_1-x_2) 
+ \int_C d^4x_3 \S(x_1,x_3)G(x_3,x_2)\;,\label{contg}\\ 
 i\slpa_1 S(x_1,x_2)&=&\d(x_1-x_2) 
+ \int_C d^4x_3 \P(x_1,x_3)S(x_3,x_2)\;,\label{conts}
\eea
where $\S$ and $\P$ are the corresponding self energies and the time 
integration is carried out along the contour $C$. Eqs.~(\ref{contg}) and 
(\ref{conts}) can be turned into matrix equations with real time integration
in the usual manner. For the off-diagonal elements $G^>$ and $S^>$ one
then obtains
\bea
C(i\slpa_1-M)G^>(x_1,x_2)&=& 
\int d^4x_3 \left(\S^>(x_1,x_3)G^-(x_3,x_2)\right. \NO\\
&&\qquad\left. +\S^+(x_1,x_3)G^>(x_3,x_2)\right)\;,\label{matrixg}\\ 
i\slpa_1 S^>(x_1,x_2)&=& 
\int d^4x_3 \left(\P^>(x_1,x_3)S^-(x_3,x_2)\right. \NO\\
&&\qquad\left. +\P^+(x_1,x_3)S^>(x_3,x_2)\right)\;.\label{matrixs}
\eea
Here $\S^>$ and $\P^>$ are off-diagonal matrix elements of $\S$ and $\P$,
which are defined analogous to $G^>$. The equations for $G^<$ and $S^<$
can be obtained from eqs.~(\ref{matrixg}) and (\ref{matrixs}) by replacing
the superscripts `$>$' by `$<$'. Equations of the type (\ref{matrixg}), 
(\ref{matrixs}) have first been obtained by Kadanoff and Baym for 
non-relativistic many-body systems \cite{kb62}. 

For processes where the overall time evolution is slow compared to relative
motions the Kadanoff-Baym equations can be solved in a derivative expansion.
One considers the Wigner transform for $G(x_1,x_2)$,
\beq
G(x,p)=\int d^4y\ e^{ipy}\ G\left(x+{y\over 2},x-{y\over 2}\right)\;,
\eeq
and $S(x,p)$, $\D(x,p)$, respectively. For the Wigner transform of a 
convolution one has in general,
\bea\label{derex}
&&\int d^4y\ e^{ipy}\ \int d^4x_2\ A(x_1,x_2) B(x_2,x_3) \\
&&= A(x,p) B(x,p) - 
{i\over 2}\left(\der x A(x,p)\der p B(x,p) - \der x B(x,p)\der p A(x,p)\right)
+\ldots\;,\NO
\eea
where $x=(x_1+x_3)/2$ and $y=(x_1-x_3)/2$.

Using the derivative expansion (\ref{derex}) the Kadanoff-Baym equations
(\ref{matrixg}) and (\ref{matrixs}) become local in the space-time coordinate
$x$. Keeping to zeroth order only the on-shell part of retarded and advanced
Green functions and self-energies, which are given by expressions analogous
to eq.~(\ref{onoff}), one obtains the equations
\bea
C({i\over 2}\slpa + \slash p - M)G^>(x,p)&=&
C({i\over 2}\slpa + \slash p - M)G^<(x,p)\NO\\
&=&{1\over 2}\left(\S^>(x,p)G^<(x,p)-\S^<(x,p)G^>(x,p)\right)\;,\label{kbN}\\
({i\over 2}\slpa + \slash k )S^>(x,k)&=&
({i\over 2}\slpa + \slash k )S^<(x,k)\NO\\
&=&{1\over 2}\left(\P^>(x,k)S^<(x,k)-\P^<(x,k)S^>(x,k)\right)\;.\label{kbl}
\eea
Solutions of these equations yield the first terms for the non-equilibrium 
Green functions $G^>(x,p)$...$S^<(x,k)$ in an expansion involving off-shell
effects and space-time variations, which include `memory effects'. As an 
example for the type of corrections we list the first derivative term on the
right-hand side of eq.~(\ref{kbN}),
\bea
\D_{\pa}&=&-{i\over 4}\left(\der x \S^>(x,p)\der p G^<(x,p) 
-\der p\S^>(x,p)\der x G^<(x,p)\right. \NO\\
&&\quad\left.- \der x \S^<(x,p)\der p G^>(x,p) 
+ \der p\S^<(x,p)\der x G^>(x,p)\right)\;. 
\eea

Solutions of the Kadanoff-Baym equations can be studied once the self-energies 
$\S$ and $\P$ are known. For weak coupling these can be determined in 
perturbation theory.\\

\noindent\textbf{Self-energies for lepton fields}\\

The one-loop contributions to the self-energy of the Majorana neutrino are 
shown in fig.~(1). For vanishing chemical potential fig.~(1a), for instance,
yields the result
\beq
-i\S^{(a)>}_{eq}(p)=2(\l^\dg\l)_{11}\int {d^4p_1\over (2\p)^4}
{d^4p_2\over (2\p)^4}(2\p)^4\d(p-p_1-p_2)C S^>(p_1)\D^>(p_2) \;.
\eeq
In the following we shall only consider the case $M\gg T$, where the heavy
neutrinos are non-relativistic. In this case the number density is small,
$f_N(E)\ll 1$, and one finds for the difference of the self-energies
($p_0>0$),
\bea
-i(\S^>_{eq}(p)-\S^<_{eq}(p))&=&-2(\l^\dg\l)_{11} \int 
d\F_{\bar{1}\bar{2}}(p) 
C \slash p_1\NO\\
&&\quad\times\left((1-f_l(p_1))(1+f_{\f}(p_2))+f_l(p_1)f_{\f}(p_2)\right)\;
\label{selfmn}\\
&\simeq&-\G C{\slash p\over M}\;.
\eea
Here $\G$ is the vacuum decay rate of the Majorana neutrino.

\begin{figure}
\input{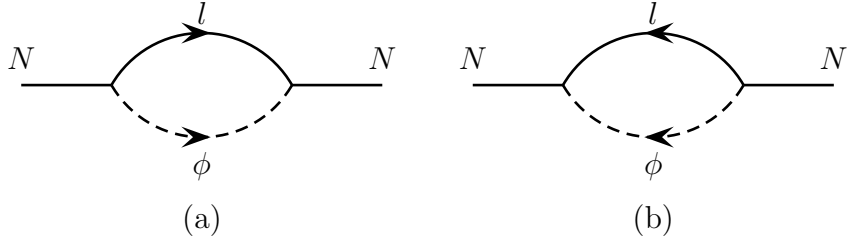}
\caption{\it One-loop self energies for the Majorana neutrino. \label{semn}}
\end{figure}

The one- and two-loop contributions to the lepton self-energy are shown in
fig.~(2a)-(2d). In the following we only list the terms which are needed for 
the solution of the Kadanoff-Baym equations to leading order. 
\begin{figure}
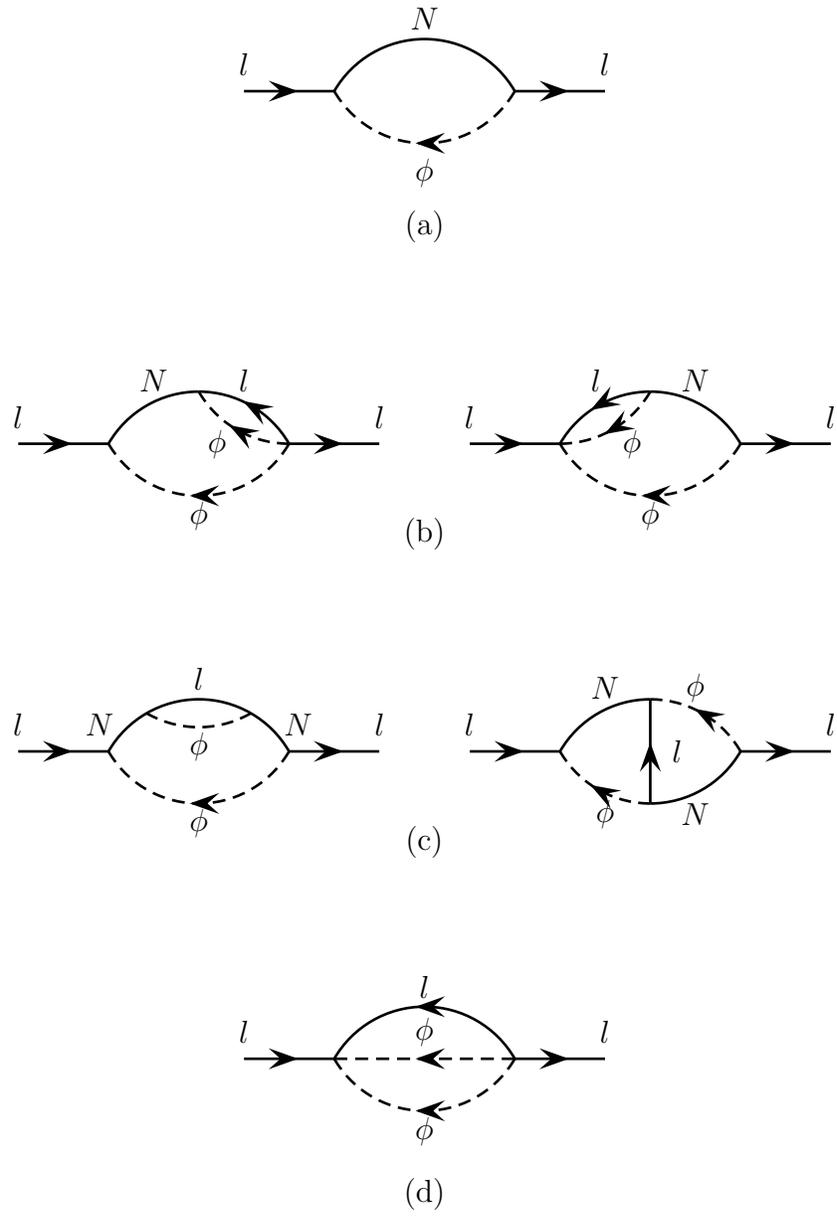

\input{fig2a.tex}
\input{fig2b.tex}
\input{fig2c.tex}
\input{fig2d.tex}
\caption{\it One- and two-loop self energies for the lepton doublet. 
\label{seld}}
\end{figure}

Particularly interesting are the terms fig.~(2b) which drive the generation
of an asymmetry. After some algebra one finds ($k_0>0$),
\beq
-i\left(\d\P^{(b)<}(k)+\d\P^{(b)>}(-k)\right) = 
{3\over 4\pi}\mbox{Im}(\l^\dg\h\l^*)_{11}M\int d\F_{1\bar{2}}(k)
\slash p_2 P_L \d f_N(t,p_2)\;.
\eeq
Here we have only given the deviation from the equilibrium self-energy
which is obtained by using for the Majorana neutrino propagator the deviation
from the equilibrium propagator $\d G \propto \d f_N$. Furthermore, we have
again considered the case of small densities. The combination
of Yukawa coulings is precisely the one occuring in the $C\!P$ asymmetry of
the Majorana neutrino decay. Since the heavy neutrinos $N_2$ and $N_3$ have
been integrated out, the contributions fig.~(2b) involve both, self-energy 
and vertex corrections \cite{fps95,crv96,bp98} which, up to a numerical 
factor, are identical in this limit.

All graphs in fig.~(2) contain washout processes. In the case of small
densities one obtains from fig.~(2a) ($k_0>0$),
\beq
-i\left(\P^{(a)>}_{eq}(k)-\P^{(a)<}_{eq}(k)\right) = -2(\l^\dg\l)_{11}
\int d\F_{1\bar{2}}(k) \slash p_2 P_L (f_{\f}(p_1)+f_N(p_2))\;.
\eeq
For $T\ll M$ the dominant contribution to the washout processes is due to
fig.~(2c) where the Majorana neutrino propagator can be replaced by a local
interaction. For small densities one finds ($k_0>0$),
\beq
-i\left(\P^{(c)>}_{eq}(k)-\P^{(c)<}_{eq}(k)\right) =  
-6{(\l^\dg\l)_{11}^2\over M^2}\int d\F_{1\bar{2}\bar{3}}(k) 
\left(\slash p_1 f_{l}(p_1) + 2\slash p_2 f_{\f}(p_1)\right)P_L\;. 
\eeq
The complete expressions for the self-energies will be given elsewhere.\\

\noindent\textbf{Kinetic equations}\\

Given the lepton self-energies we can now look for solutions of the 
Kadanoff-Baym equations (\ref{kbN}) and (\ref{kbl}). A straightforward
calculation shows that the right-hand side of these equations vanishes
for equilibrium Green functions and self-energies. Since baryogenesis is
a process close to thermal equilibrium we can search for solutions which
are linear in the deviations,
\bea
\d G(t,p) &=& G^>(t,p)-G^>_{eq}(p) = G^<(t,p)-G^<_{eq}(p)\;,\\
\d S(t,p) &=& S^>(t,p)-S^>_{eq}(p) = S^<(t,p)-S^<_{eq}(p)\;.
\eea
One then obtains for the perturbations $\d G(t,p)$ and $\d S(t,p)$,
\bea
iC\g^0\der t \d G(t,p) &=& iC\g^0\der t G_{eq}^>(p)  
 +(\S^>_{eq}(p)-\S^<_{eq}(p))\d G(t,p)\;, \label{kbdn}\\
i\g^0\der t \d S(t,k) &=& (\P^>_{eq}(k)-\P^<_{eq}(k))\d S(t,k)\NO\\
&& + \d\P^>(t,k) S^<_{eq}(k) - \d\P^<(t,k) S^>_{eq}(k)\;.\label{kbds}
\eea
The Green functions depend on time explicitly, as well as implicitly
through the time-dependence of the temperature. Once the `covariant' 
time derivative is used, the later vanishes for equilibrium Green functions
of massless fields. This is not the case for massive fields. Hence, the first
term on the right-hand side of (\ref{kbdn}) drives the deviation from
thermal equilibrium.

We can now insert the perturbative expressions for the self-energies into
eqs.~(\ref{kbdn}), (\ref{kbds}) and check whether the ansatz (\ref{devn}),
(\ref{devl}) for $\d G$ and $\d S$ yields a solution. After some algebra
one finds that this is indeed the case provided the distribution functions
$\d f_N$ and $\d f_l$ satisfy the following ordinary differential equations,
\bea
E\der t \d f_N(t,p) &=& 
-E\der t f_N(p) - 2 (\l^\dg\l)_{11} \int d\F_{\bar{1}\bar{2}}(p)
\d f_N(t,p) p\cdot p_1\;,\label{diffn}\\
g_l k \der t \d f_l(t,k) &=&  
{3\over 8\p}\mbox{Im}(\l^\dg\h\l^*)_{11}M \int d\F_{\bar{1}2}(k)
\d f_N(t,p_1) k\cdot p_1\ \NO\\
&&-2 (\l^\dg\l)_{11}\int d\F_{1\bar{2}}(k) 
\left(\d f_l(t,k)f_\f(p_1)+f_l(k)\d f_\f(t,p_1)\right) k\cdot p_2 \NO\\
&&-6 {(\l^\dg\l)^2_{11}\over M^2} \int d\F_{1\bar{2}\bar{3}}(k)\Big(
2(\d f_l(t,k)f_\f(p_1)+f_l(k)\d f_\f(t,p_1)\NO\\
&&\hspace{2cm}+\d f_l(t,p_2)f_\f(p_3)+f_l(p_2)\d f_\f(t,p_3))
k\cdot p_2 \NO\\
&&\hspace{2cm}+(\d f_l(t,k)f_l(p_1)+f_l(k)\d f_l(t,p_1)\NO\\
&&\hspace{2.5cm}+2\d f_\f(t,p_2)f_\f(p_3)) k\cdot p_1\Big)\ .\label{diffl}
\eea
Comparing these equations with the Boltzmann equations (\ref{bon}) and 
(\ref{bol}) one finds that the two sets of equations are identical
to leading order in the coupling where matrix elements and $C\!P$ asymmetry 
are given by
\bea
|\cm(N(p)\rightarrow l(p_1)\f(p_2))|^2 &=& 4 (\l^\dg\l)_{11}\ p\cdot p_1 \ ,\\
|\cm(l(k)\f(p_1)\rightarrow \bar{l}(p_2)\bar{\f}(p_3))|^2 &=& 
24 \frac{(\l^\dg\l)^2_{11}}{M^2}\ k\cdot p_2\;,
\eea
\beq
\e = {3\over 16\p}{\mbox{Im}(\l^\dg\h\l^*)_{11}\over (\l^\dg\l)_{11}}M\ .\\
\eeq
We conclude that for non-relativistic
heavy neutrinos a solution of the Boltzmann equations generates a
solution of the full Kadanoff-Baym equations to leading order in the
expansion described above. For relativistic heavy neutrinos the matrix
structure of the equations is more complicated and the time evolution of
the different poles of the Majorana neutrino propagator are described 
by different equations.

Given a solution of the Kadanoff-Baym equations to leading order the various
corrections can be systematically studied. Note, that the solutions of 
eqs.~(\ref{diffn}) and (\ref{diffl}) are not of the form 
$\d f_i(t,p) = h_i(t) f_i(p)$. Hence, the usual assumption of kinetic
equilibrium does not appear to be justified. The size of `derivative terms',
which correspond to memory effects, and off-shell corrections can be
determined by inserting the leading order solution into the various correction
terms described above. Particularly interesting are relativistic corrections 
in the case that leptogenesis takes place at temperatures $T\sim M$. 

The analysis of the Kadanoff-Baym equations for leptogenesis can be used to
obtain constraints on the parameters $M$, $(\l^\dg\l)_{11}$ and $\e$, 
which provides a quantitative relation between the cosmological baryon 
asymmetry and neutrino properties.

\end{document}